\documentclass[12pt]{article}
\usepackage{hyperref}

\usepackage{float}
\usepackage{amsmath}
\usepackage{amssymb}
\DeclareSymbolFontAlphabet{\mathcal}   {symbols}
\usepackage{graphicx}
\usepackage{algorithm}
\usepackage[noend]{algpseudocode}
\usepackage{caption}
\usepackage{subcaption}
\usepackage{multirow}
\usepackage{hhline}
\usepackage[T1]{fontenc}
\usepackage[utf8]{inputenc}
\usepackage[numbers]{natbib}
\usepackage{multirow}
\usepackage{xcolor}

\begin{document}
\hypersetup{pdfauthor={Brendan Cross}}

\title{Hierarchical community detection benchmark for heterogeneous inter-community connectivity}

\author
{
Brendan Cross$^{\text{1}}$, Boleslaw K. Szymanski$^{\text{1,*}}$ \\
\normalsize{$^{\text{1}}$Department of Computer Science and Network Science and Technology Center,}\\
\normalsize{Rensselaer Polytechnic Institute, Troy, NY, USA}\\
\normalsize{$^{*}$Corresponding author, email: szymab@rpi.edu}
}

\maketitle 

\begin{abstract}
Here, we introduce a new tool for community detection, a generator of networks, which uses parameters to control the structure of created networks. Typically, network scientists designing novel community detection algorithms use synthetically generated benchmarks with community structures that they intend to detect and scale the benchmark networks across size and density. Currently, available benchmarks use generators limited to the properties of the LFR and GLFR networks. We improve on these previous benchmarks with a new hierarchical benchmark, the HGLFR, that preserves the properties of the LFR and GLFR while extending them to include heterogeneous inter-community connectivity. Networks generated by this benchmark are shown to produce networks with structures triggering the resolution limit while maintaining assortative connectivity.
\end{abstract}



\section{Introduction}
    Networks are structures that reveal how basic units of a system interact or relate to one another. Network scientists construct these networks to represent real-world systems. Analyzing these networks is essential to understanding the represented system and its components. An important property of networks is that groups of nodes connect to other groups of nodes to form modules or communities. The analysis of these communities and how they form in real-world systems are at the core of network science. For example, such analyses have proven to be effective in studies of polarization of users in social networks~\cite{soares2018influencers, flamino2023political}, in searching interconnected functional groups of neurons in the brain~\cite{brooks2024community}, or in understanding processes in other biological systems~\cite{sporns2016modular, berahmand2021spectral}. Thus, it is essential to have robust methods for identifying community structures in these networks.
    
    In these networks, community structures can form complex hierarchies, where groups of communities may favor connections with each other at the expense of different groups of communities within the network. These structures are present in many systems, such as political leanings in social networks. Another example may be the clustering of interaction networks, where a student may have dense connectivity to people within the same school. Within the school, we could find a more dense clustering of students who share the same major or the same classes.
    
    The development of new community detection methods relies on the availability of robust test data. We need access networks with known community structures to evaluate any process's performance. To ensure that the performance of the methods is generalizable to real-world networks, we need to generate networks similar to those we intend to detect communities. Community detection benchmark generators provide us with both of these properties. These generators can model their structural properties based on the real-world networks they seek to emulate. Once they do so, these generators can create large sets of networks of any size required, all with predicted, known community structure.
    
    In this paper, we introduce the HGLFR benchmark generator, an extension to the widely used Lancichinetti-Fortunato-Radicchi (LFR) benchmark generator ~\cite{lancichinetti2008benchmark}. This extension introduces a hierarchical community structure to the LFR benchmark, allowing the inter-community connectivity to vary based on the sampled hierarchical structure. We show how this inter-community heterogeneity is directly related to the asymptotic resolution bounds of generalized modularity~\cite{lu2020asymptotic} and how this heterogeneity impacts the performance of different community detection methods.

\section{Related Work}

    \subsection{Community Detection Benchmarks}

        Community detection benchmarks are synthetic random networks with incepted community structure, that are meant to be robust enough to validate community detection algorithms on. The most important tasks of community detection benchmark networks are that they have known, well-formed, community structure and that the networks are generated with properties that mirror real-world networks (either generally or targeting specific network types). The LFR benchmark network~\cite{lancichinetti2008benchmark} is the most often used method for generating artificial networks with real-world properties. This benchmark is able to achieve our first goal in that it reliably generates known community structures in its networks, and has many desirable properties relating to the second goal. LFR networks contain heterogeneous node degree and community size distributions, allowing the networks to mirror the variation seen in real world systems. 
        
        Despite these desirable properties, it has been shown that the community structures generated by this benchmark are homogeneous across other network properties~\cite{le2017glfr}. Specifically, the LFR network is shown to be homogeneous in the mixing parameter for each community, leading to all communities in the network sharing a very similar internal to external degree ratio. To address this issue, the authors propose a more general form of the LFR benchmark, the GLFR benchmark~\cite{le2017glfr}, that generates communities with heterogeneous mixing parameters, better reflecting the performance of algorithms on real work networks, compared to LFR. The author's accomplish this by sampling the mu for each community uniformly in the range of $[\mu - \Delta_\mu, \mu + \Delta_\mu]$.

        There has also been work in extending the LFR benchmark to generate hierarchical community structure. In work by Yang et al.~\cite{yang2017hierarchical}, the authors combine the LFR benchmark's desirable heterogeneous properties with the Rabasz-Barabasi method of hierarchical network generation~\cite{ravasz2003hierarchical}. Additionally, the authors provide methods of removing excessive replication within the network by introducing a parameter to control for the decimation of replicated communities.

    \subsection{Resolution Limit and inter-connectivity heterogeneity}

        Heterogeneity in various network properties can cause a host of issues in community detection methods. One such issue is known as the resolution limit of generalized modularity~\cite{fortunato2007resolution, lancichinetti2011limits}. This issue, within the class of modularity based community detection methods~\cite{newman2006modularity}, occurs when there is no value for the resolution parameter that is able to recover the desired partition of a network.

        Generalized modularity~\cite{reichardt2006statistical}, defined by equation \ref{eq:modularity}, is a metric describing how modular a given network partition is. In this equation, $A$ is the adjacency matrix of the network, $k_i$ is the degree of node i, $m$ is the number of edges in the network, and $\delta(c_i, c_j)$ is 1 if the communities of i and j are the same and 0 otherwise. The metric defines a resolution parameter $\gamma$ that allows us to adjust how coarse or fine grained the discovered community structure is. With the resolution parameter set to 0 the subtraction term would become 0, making optimal a network partition that includes every node in the network in one large community. Likewise, an arbitrarily large resolution penalizes internal edges in a community so much that the highest modularity partition is a singleton network, where each node becomes its own community.
        \begin{equation}
            Q = \frac{1}{2m}\sum_{ij}\Big[ A_{ij} - \gamma\frac{k_ik_j}{2m} \Big] \delta(c_i, c_j)\label{eq:modularity}
        \end{equation}

        In work by Lu et al.~\cite{lu2020asymptotic} a range of values for $\gamma$ is defined that can be used to recover any target community structure using modularity maximization methods. This range is defined as a function of a partition's $\Omega$ matrix, which is a $C\times C$ matrix where $C$ is the number of communities in the partition and each entry $\Omega_{i,j}$ defines the ratio of actual to expected edges between communities $i$ and $j$. The null model, which defines the expected edges between communities, is configuration model~\cite{newman2003structure}, a network generator that creates a random network with prescribed degree sequence. In the $\Omega$ matrix, the resolution bounds are determined by the minimum intra-community density $min(\Omega_{i,i})$ and the maximum inter-community density $max(\Omega_{i,j})$. 

\begin{figure}[!htbp]
            \hspace*{-0.9cm}
            \centering
            \begin{subfigure}{0.6\textwidth}
            \includegraphics[width=1.0\linewidth]{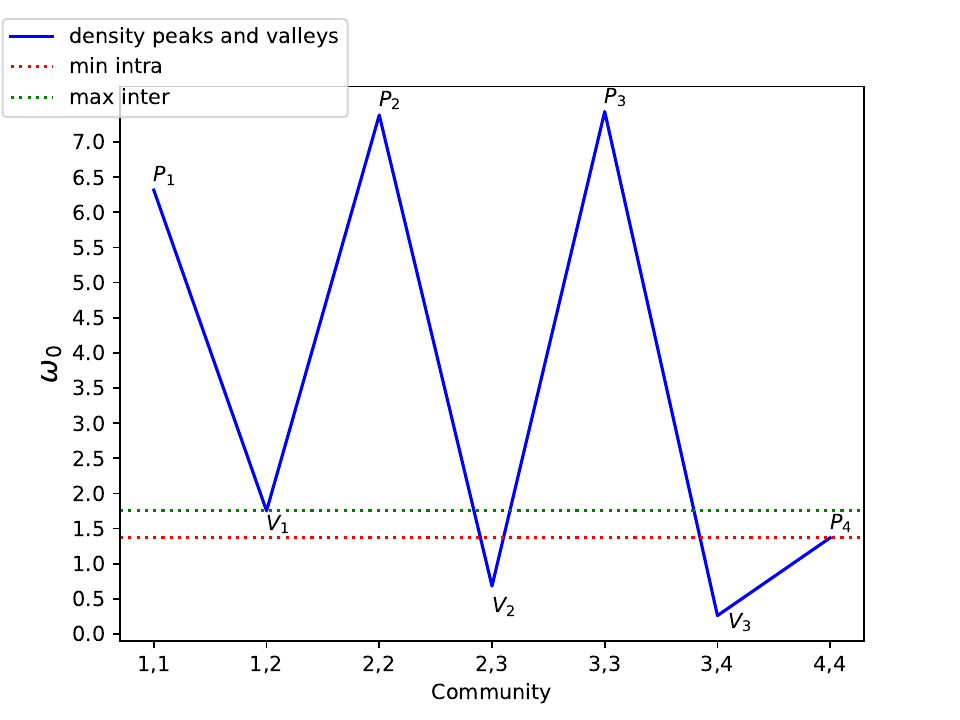}
            \caption{Peaks and valleys}
            \label{fig:peak_subplot}
            \end{subfigure}
            \hspace*{-0.7cm}
            \begin{subfigure}{0.4\textwidth}
            $D = min(\Omega_{i,i}) - max(\Omega_{i,j})$
            \begin{equation}
            \Omega = 
            \left[
            \begin{tabular}{c c c c}
            \textcolor{blue}{6.31} & \textcolor{red}{1.75} &0.15 &0.13 \\
            \textcolor{red}{1.75} &\textcolor{blue}{7.38} &\textcolor{black}{0.68} &0.08 \\ 
             0.15 &\textcolor{red}{0.68} &\textcolor{blue}{7.43} &\textcolor{black}{0.26}\\
             0.13 &0.08 &\textcolor{red}{0.26} &\textcolor{blue}{1.36}\\
             \end{tabular}
             \right]\nonumber
             \end{equation}
             \caption{Community density matrix}
             \end{subfigure}
             \caption\textbf{{Community Peaks and Valleys.} Each entry along the x-axis corresponds to $i,j^{th}$ entry of the community density matrix ($\Omega$), which happens to correspond to the peak of each community, highlighted in blue, followed by the largest intra-community density between adjacent peaks, highlighted in red.}
            \label{fig:ToyExample_PeakValley}
        \end{figure}

        Here, we quantify how heterogeneous a network is through the use of these asymptotic resolution bounds. We define the term $D$ to be the distance of this resolution window, where it is positive when there is a range of resolutions that can recover the true partition and negative when there are no values. Figure \ref{fig:ToyExample_PeakValley} defines this term as a function of the $\Omega$. This distance metric $D$ 
        
        To illustrate this, let's consider an undirected network containing 250 nodes and four well defined communities. In Figure\ref{fig:ToyExample_PeakValley}, we represent the densities as four peaks and three valleys. The peaks represent the densities of edges internal to each of the four communities of the network while the valleys represent the largest densities of edges between pairs of communities. The  figure uses the green and red horizontal lines to represent our resolution bounds.

        A key point about these resolution bounds, as it relates to this paper, is what causes these resolution window endpoints. $min(\Omega_{i,i})$ represents the weakest connected community, in terms of density, in the network. This value is decreased when either a community grows large and encompasses much of the expected degree of a network and when the community is connected weakly when compared to other communities in the network. $max(\Omega_{i,j})$ on the other hand is the stronger connective density between a pair of communities. This value is maximized in the case of a pair, or group, of communities having are densely connected to each other and as those communities decrease in size. When this inter-community connectivity exceeds the weakest intra-community connectivity we express the resolution limit. This inter-community density is the key to hierarchical heterogeneity and its relation to our distance metric $D$. As we have a stronger hierarchical component in our network, the lower bound of our resolution window increases, decreasing $D$. When $D$ becomes negative, we have structures that exist at different resolutions / scales that are strongly connected and competing. We define this heterogeneity as heterogeneous inter-community connectivity.

        In work by Lu et al.~\cite{lu2020asymptotic}, the authors define a range of values for $\gamma$ that recover any target community structure using modularity maximization methods. This range is a function of a partition's $\Omega$ matrix, which is a $C\times C$ matrix where $C$ is the number of communities in the partition and each entry $\Omega_{i,j}$ defines the ratio of actual to expected edges between communities $i$ and $j$. The null model, which describes the expected edges between communities, is configuration model~\cite{newman2003structure}, a network generator that creates a random network with a prescribed degree sequence. For this $\Omega$ matrix, the resolution bounds are the minimum intra-community density $min(\Omega_{i,i})$ and the maximum inter-community density $max(\Omega_{i,j})$. 

\section{Hierarchical Generalized LFR}

    Here, we outline a hierarchical benchmark for community detection that builds on the heterogeneous properties of the LFR and GLFR benchmarks while introducing hierarchical community structures to accomplish two things: the first adds an inter-community heterogeneity to the network, while the second allows for the natural modeling of networks with the resolution limit, overcoming the course of resolution limit, while still using the classic modularity based algorithm for community detection.

    We define inter-community heterogeneity as how the selection of a node for external edge varies over community membership in a network. Networks with low inter-community heterogeneity would be expected to have no hierarchical clustering, with each community and node sampling external edges uniformly from all external nodes. With high inter-community heterogeneity, we expect a very hierarchical network, with communities preferentially attaching to groups of other communities, repeating for several levels of generalization. This dynamic makes the community detection landscape more complex since, in addition to the ground truth communities being clusters defined by modularity, groups of communities themselves will also have modular structures with each other.

    In the LFR generator, all external connections from a community are sampled uniformly across all other nodes in the network. This uniformity provides a predictable and, importantly, homogeneous property for all nodes in the network. In LFR networks, each node has $(1-\mu)\times k$ internal connections and $\mu \times k$ external connections. 
    The GLFR model allows the $\mu$ to vary per community uniformly within the range $[\mu - \delta, \mu + \delta]$. Still, we retain the property of uniformly selecting external edges from all non-community nodes. 
    When the connectivity between nodes of different communities is regular, this uniformity makes community detection easier than in many real-world networks, whose communities may preferentially favor connections to specific other communities or groups of nodes. 
    In work by Peel et al.\cite{peel2018multiscale} they highlight this behavior in multiple real world networks, which are shown to contain wide varieties of structures with heterogeneous mixing patterns.

    In addition to better representing the properties of real-world networks, inter-community heterogeneity is also related to the resolution limit of generalized modularity. As described previously, we can define a range of resolution values for any given network and desired partition, which the algorithm can use with modularity maximization to recover the target partition. The endpoints of this resolution range are the smallest intra-community connection density ($min(\Omega_{i, i})$) and the largest inter-community connection density ($max(\Omega_{i,j})$). Knowing these densities, we can compare the actual versus expected number of edges between nodes within and between communities. For the resolution limit described in~\cite{lu2020asymptotic} to occur, we need $max(\Omega_{i,j})$ to be larger than $min(\Omega_{i,i})$. For this property to arise, we need groups of communities to be more connected to each other than we would expect from uniformly sampling external edges. To enforce such a dynamic, we need hierarchical groupings of communities. We show that adding such a hierarchical community structure increases the ability of the benchmark to generate networks with this resolution limit issue.

    The HGLFR generator introduces an arbitrary number of hierarchical groupings of the generated ground-truth communities. We define four parameters of this hierarchy. The first, $L$, determines the number of levels within this hierarchy. The second parameter $\mu_L$ denotes the mixing parameter at each level of the hierarchy. The third parameter $\Delta_{\mu_L}$ denotes the hierarchical mixing deltas for each level of the hierarchy. The final parameter is $S$, which defines the probability of communities merging at each hierarchical level when generating the community hierarchy. In Algorithm \ref{algo:1}, we describe the process of generating an HGLFR network.

    \begin{algorithm}[H]
    \small
    \caption{Hierarchical LFR}
        \begin{algorithmic}[1] 
        \State Generate the degree distribution by sampling \textit{N} values from the power-law distribution defined by the three parameters <$k$>, $k_{max}$, $\tau_1$.
        
        \State Generate the community size distribution by sampling $c_{min}$, $c_{max}$, $\tau_2$
        
        \State Sample an $L$ level hierarchical clustering. Starting with $l_0$, the base communities defined by step 2 generate $l_1$ by randomly merging communities of $l_0$ with probability $S$, ensuring at least one merge occurs. Repeat until we have $L-1$ levels. The last level will be the entire network as one community.
        
        \State Determine the mixing parameter for each community to each other community using the hierarchical mixing parameter $\mu_L$, the hierarchical delta $\Delta_{\mu_L}$, and the hierarchical clustering from step 3. We sample the mixing parameter to any community $c$ from the range $[\mu_L^i-\Delta_\mu^i, \mu_L^i+\Delta_{\mu}^i]$ where $i$ is the level the communities are first members of the same hierarchical community. 
        
        \State Assign all internal edges for each community using the configuration model to keep the community's internal degree distribution. 
        
        \State Assign all external edges for each node weighted by the mixing parameter of each hierarchical level defined in step 4. 
        
        \end{algorithmic}
        \label{algo:1}
    \end{algorithm}

    The first step is to generate the degree distribution for all nodes in our network. We create samples from a target power-law distribution defined by a given average degree <$k$>, max degree $k_{max}$, and negative power-law coefficient $\tau_1$. The second step performs a similar sampling for the community size distribution constrained by the minimum community size $c_{min}$, maximum community size $c_{max}$, and negative power-law coefficient $\tau_2$. The HGLFR algorithm utilizes the same approach for sampling these distributions outlined in the original LFR model.

    The third step is to generate an $L$ level hierarchy using two new parameters: $L$ and $S$. $L$ determines the number of levels in the final hierarchy, and $S$ indicates the probability of each community merging with another community as we generate the hierarchy. This $S$ parameter acts as an inter-connectivity parameter, where a value closer to 1 will create a hierarchy where more communities group together more quickly, and a value closer to zero will have fewer communities grouping in the other hierarchical levels. In this step, we must ensure that each successive level of the hierarchy has fewer communities than the previous level to ensure we have $L$ unique levels.

    The fourth step is to assign nodes to our base-level communities and then assign those communities to the hierarchical groups defined in Step 3. Nodes are assigned to communities randomly but with a constraint to ensure good behavior. This constraint is that the required internal edges must be less than the size of the community: $k_n \times (1-\mu_c) < |c| - 1$, ensuring that the node can assign all internal edges. We next assign a mixing parameter, $\mu$, to each community in the network. The value of this parameter is varied at each level based on the corresponding hierarchical mixing parameters $\mu_L$ and $\Delta_\mu$. These parameters are both $L$ length vectors, defining the mixing fraction of external edges at each level and how much these mixing fractions are allowed to vary per community. A community's $\mu$ to a particular level of the hierarchy, $i$, is either sampled from a range of $[\mu_L^i-\Delta_\mu^i,\ \mu_L^i+\Delta_\mu^i]$ if the community belongs to a new group at level $i$ or it is zero otherwise. Each community will have at least $\mu_{L-1}$ value if it belongs to no hierarchical groups. The expected external degree of each community to each level of the hierarchy is $\mu_L^i * k_c$, and the expected internal degree is $(1 - \sum_{i=0}^L{\mu_L^i}) * k_c$.
    
    Step 5 and step 6 involve assigning the edges for each node in the network. We compute the internal and external degrees for each community as described in step 4. We then randomly sample internal edges using the configuration model, matching the internal degree sequence for the community defined in step 4. After the internal edges are assigned, the remaining edges necessary for each node are assigned to external communities based on the generated hierarchy. For each community $c$ in the network we loop over each level of the hierarchy $l$. At each level, we determine if community $c$ is connected to any new communities at this level, if it is we form external links to the set of those new communities with probability proportional to $c$'s mixing parameter at this level $\mu_l^i$.
    
    For example, if we have three ground truth communities $A, B,$ and $C$ with a hierarchy of $L_1 = [\{A,B\}, \{C\}]$ and $L_2 = [\{A,B,C\}]$, we would first form all the internal edges for each community $A, B,$ and $C$. Next, we would next connect communities $A$ and $B$ together with the $\mu$ value assigned to level $L_1$ from our input parameters ($\mu_1$). Finally, we connect community $A$ to $C$ with the $\mu_2$ and community $B$ to $C$ with that same $\mu_2$. Since community $C$ is added to the $A, B$ cluster at $L_2$, $A$ and $B$ need to assign external edges to $C$ based on the $\mu_2$ parameter.
    
    Assigning edges optimally is computationally expensive, and often, the resulting degree distribution for nodes may be slightly off due to the constraints of assigning particular fractions of edges to each level and obeying the target degree distribution. In our implementation, we allow minor discrepancies in the target degree distribution to allow for more consistent network generation.

    \begin{figure}[!hbp]
        \centering
            \includegraphics[width=.8\textwidth]{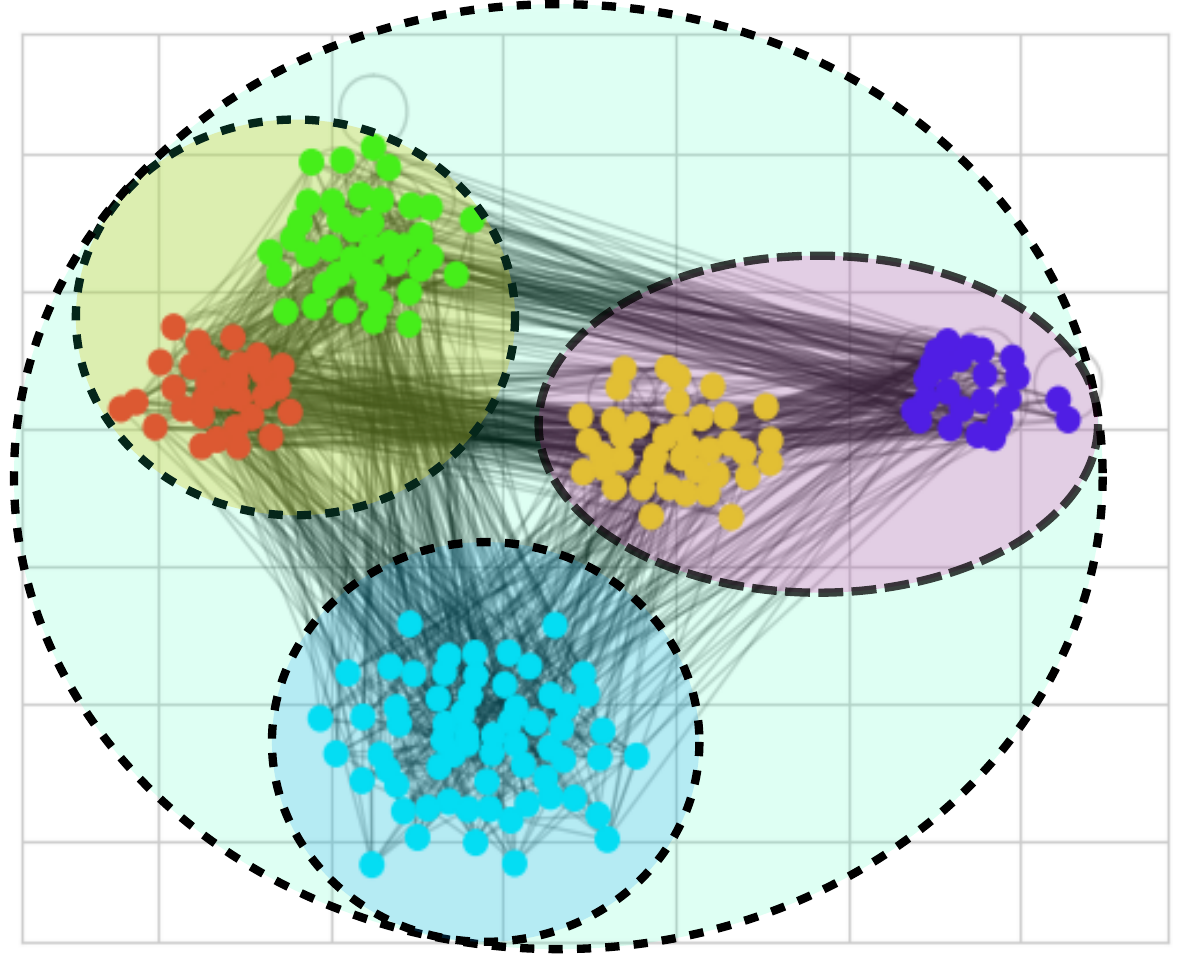}
        \caption{\textbf{Sampled community hierarchy.} In this figure we highlight the how the levels of the hierarchy are generated. We begin with a set of communities comprised of nodes with prescribed degree, but with no connections assigned. For each community we calculate the required external degree to each level of the hierarchy to match the level connectivity parameter. We then perform a random merging step, beginning from the ground truth communities. Looping over each community we merge it to an existing community with probability S. After groupings are assigned, we perform a shuffling step where communities are switched to hierarchical groups based on their ability to satisfy their necessary external connectivity.}
        \label{fig:network_diagram}
    \end{figure}

   \section{Validation}
    We evaluate and compare the HGLFR model to the LFR and GLFR benchmarks. This step aims to establish that the networks generated by the HGLFR operate as intended and that the generated hierarchies are detectable. To do this, we will first generate a test set of networks with each of the LFR, GLFR, and HGLFR generators. Where possible, we will use the same parametrization for each generated network. Since the GLFR and HGLFR networks are additive in parameters, parameters are shared and set to the same values between the different models. With the test networks, we next validate that the properties of each generated network are similar where expected and analyze the differences where they occur.

    The second step of validation is to evaluate the generated community structure. Utilizing various community detection methods, we will analyze how these methods perform on the HGLFR networks compared with LFR and GLFR and confirm, through modularity maximization methods at different resolutions, that the hierarchical community structure is detectable and behaves as expected.

    \subsection{Generated Network Properties}
\begin{figure}[H]
            \centering
                \includegraphics[width=.7\textwidth]{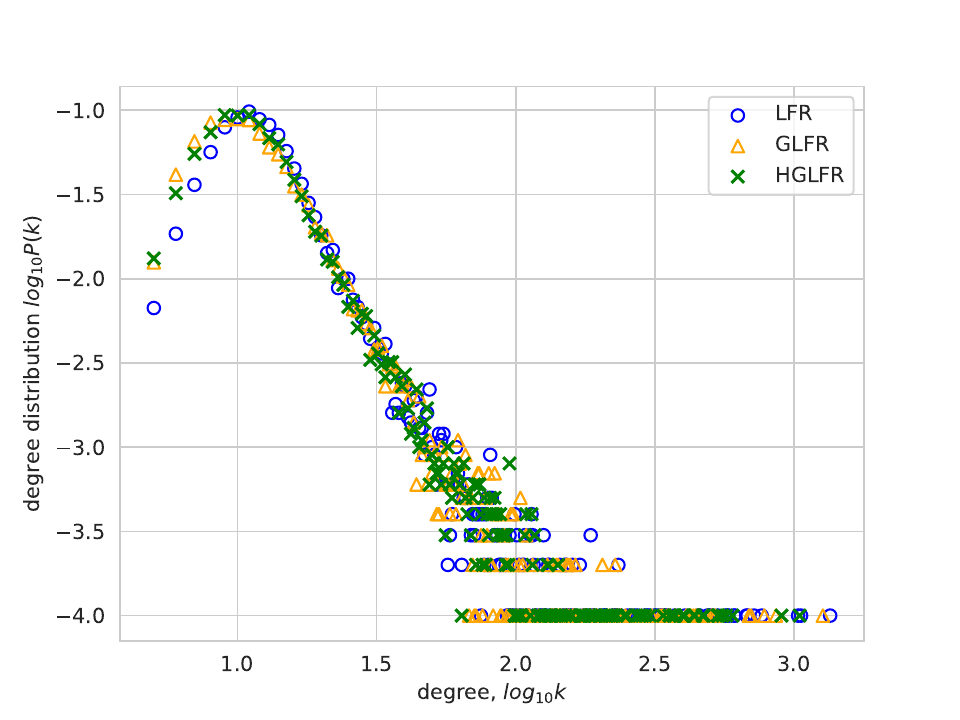}
            \caption{\textbf{Generator degree sequences}. The LFR, GLFR, and HGLFR generators generate degree distributions with reference sampled power law.}
            \label{fig:deg_seq}
        \end{figure}

        First, we compare the baseline properties of the networks generated by the HGLFR generator to the LFR and GLFR benchmarks. Figure \ref{fig:deg_seq} compares the generated degree distributions of the LFR, GLFR, and HGLFR networks to the raw power-law samples for the same parametrization. This figure shows that the networks generated by the HGLFR closely follow those generated by the non-hierarchical LFR and GLFR benchmarks. The HGLFR generator samples its degree and community size distributions in the same way as the LFR and GLFR networks, and is able to preserve these distributions during the hierarchy formation and edge connection steps.
        
        We note here that the degree distribution for each of the network generators has a dip in the probability of low-degree nodes; this is an implementation-specific result, as the HGLFR extends the python package Networkx's~\cite{hagberg2008exploring} implementation of the LFR benchmark.
    
\begin{figure}[H]
            \centering
            \begin{subfigure}[t]{0.78\textwidth}
                \centering
                \includegraphics[width=1\linewidth]{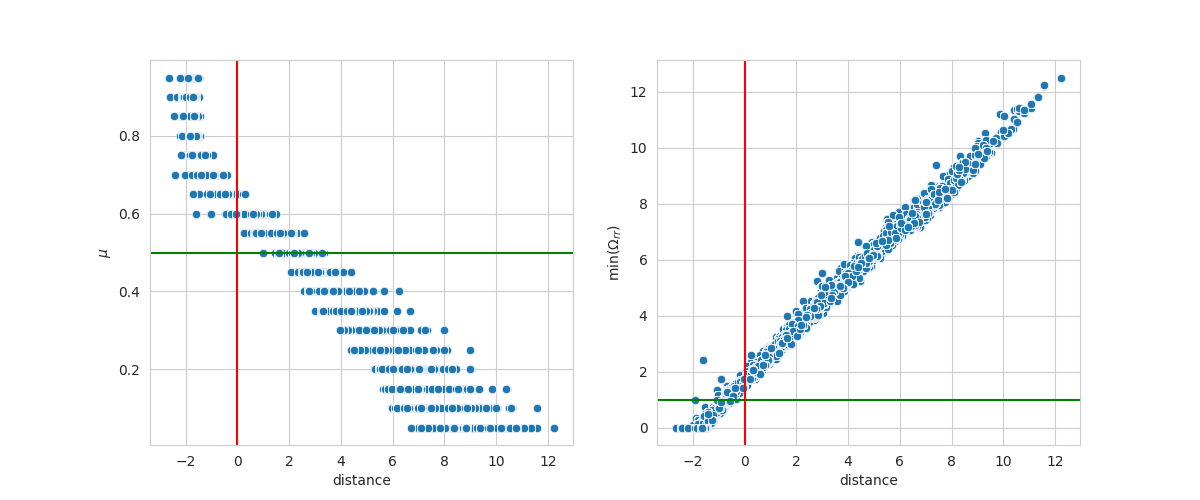}
                \caption{LFR}
                \label{fig:generated_distance_LFR}
            \end{subfigure}
            \\
            \begin{subfigure}[t]{0.78\textwidth}
                \centering
                \includegraphics[width=1\linewidth]{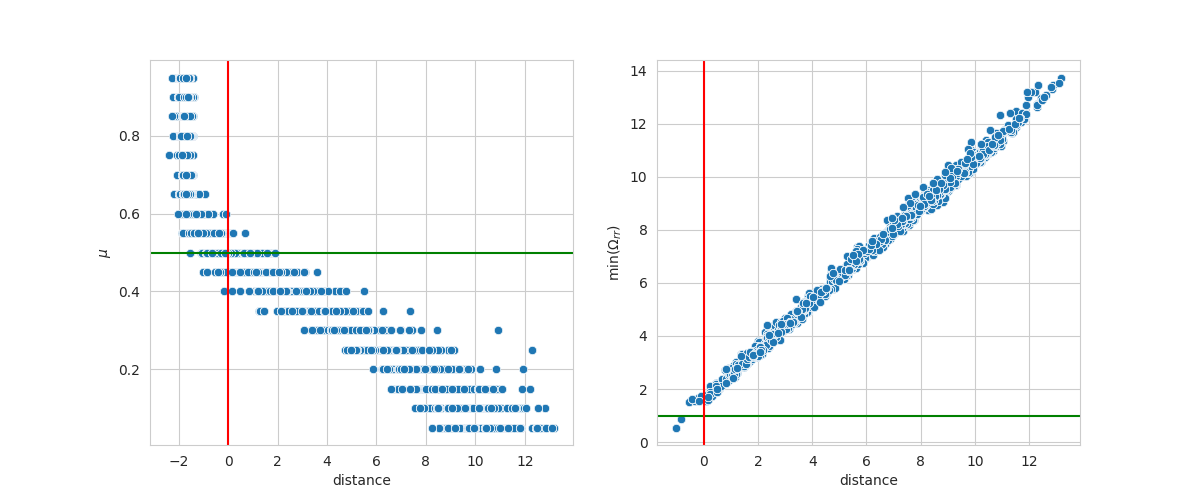}
                \caption{GLFR}
                \label{fig:generated_distance_GLFR}
            \end{subfigure}
            \\
            \begin{subfigure}[t]{0.78\textwidth}
                \centering
                \includegraphics[width=1\linewidth]{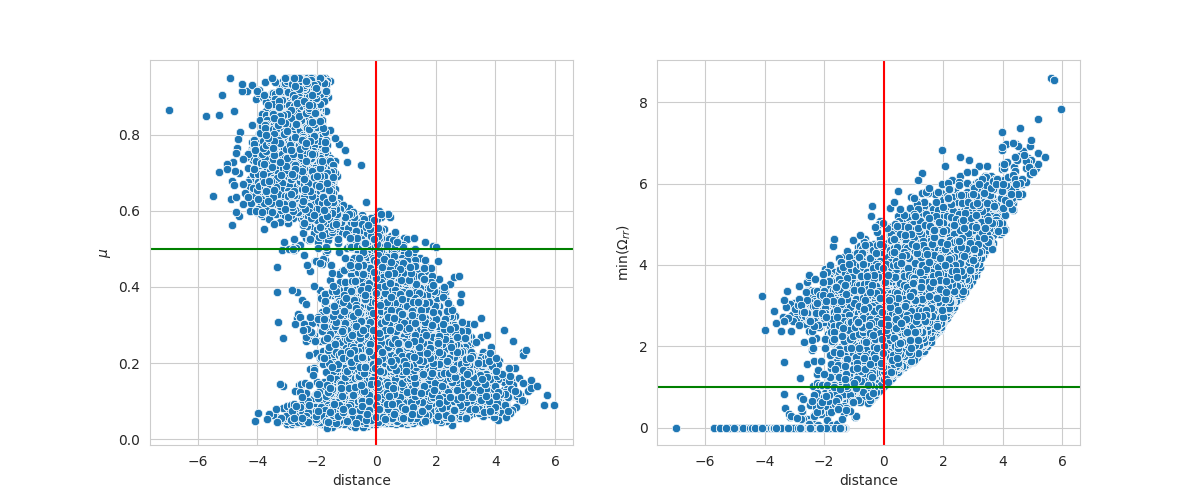}
                \caption{HGLFR}
                \label{fig:generated_distance_HGLFR}
            \end{subfigure}
    
            \caption{\textbf{Generated networks by resolution window distance.} This figure shows the resolution window distance, $D$, for networks generated by the LFR (a), the GLFR (b) and the HGLFR (c) as a function $\mu$ and the minimum intra-community connectivity $min(\Omega_{i,i})$. The green line indicates the boundary of assortative network communities ($\mu = 0.5$ or $\Omega_{i,i} < 1$). The red line indicates the boundary of negative distance.}
            \label{fig:generated_distance}
        \end{figure}

        Next, we compare the inter-community heterogeneity produced by each generator. We measure this heterogeneity by $D$, the magnitude or range of valid resolution values for the target ground-truth communities. A range of resolutions can recover the ground truth when this distance is positive. When the distance is negative, no resolution will avoid splitting a community or merging multiple communities. A network with high inter-community heterogeneity will have a negative distance because of the high fraction of inter-community edges between hierarchical groups. Figure \ref{fig:generated_distance} shows the $D$ of the generated networks as a function of the provided mixing parameter    $\mu$ for the left subplot and the minimum intra-community density $min(\Omega_{r,r})$ for the right subplot. 
        
        We expect two cases will produce networks with negative distance. The first case naturally occurs as a function of the mixing parameter $\mu$. When $\mu > .5$ the generated networks become dis-assortative, where communities have less edges internally than externally. Since our distance is defined as $D = min(\Omega_{i,i}) - max(\Omega_{i,j})$, we would expect that distance naturally trends towards negative values in dis-assortative networks. These cases are not interesting for our purposes, since we care to generate networks that contain the resolution limit in assortative networks. The second case in which we have negative $D$ is when we have some dense hierarchical connectivity in an assortative network. 
        We can see, in the left column of subplots in Figure \ref{fig:generated_distance} $D$ as a function of the mixing parameter $\mu$ for each network generator. The red horizontal line indicates the separation of positive and negative resolution window distances $D$. The green horizontal line indicates $\mu = .5$, a rough boundary for assortativity. We expect networks with $\mu > 0.5$ will begin to have negative distance without containing the resolution limit because dis-assortative networks, by definition, have less than random edges within communities with more forming externally. Networks with $D < 0$ indicate networks that have areas of dense inter-community connectivity and networks that have a $\mu < .5$ are likely to be assortative. Networks with inter-community heterogeneity exist in the space of $\mu < .5$ and $D < 0$.  To reliably generate networks with this resolution limit, we should generate negative distance networks for a wide range of $\mu$.

        Comparing the performance of the generators, we see that for the LFR generator, we produce $D<0$ networks only in the case of dis-assortative networks, where $\mu > .5$. The reason is that all external edges are sampled uniformly across all non-community nodes, so the likelihood of dense preferential connectivity between two communities in the network is low. The GLFR generator can process a minimal number of networks, and the range of negative distances is restricted. The HGLFR generator, on the other hand, can generate a significantly wider variety of networks across the entire range of positive and negative distances. These results show that the HGLFR can generate networks containing the resolution limit described in~\cite{lu2020asymptotic} and create networks with a wide degree of inter-community heterogeneity.
        
    \subsection{Algorithm Performance}
        In this section, we examine the performance of different community detection methods on the new HGLFR generator and its predecessors, LFR and GLFR. For the comparison, we generated a large set of networks varying the mixing parameters of the networks. Table \ref{tab:BaseLFRs} defines the exact parameters used and lists the shared parameters between LFR, GLFR, and HGLFR in the left table. The HGLFR differs from the LFR and GLFR in that it defines a $\mu$ and $\Delta$ parameter for each hierarchy level. To generate the test network set for HGLFR, we generated networks under three hierarchical parametrizations, computed the $\mu$ achieved for the entire network, and binned them into $[.05, .1, \hdots, .9, .95]$ bins. These three parameterizations are defined in the right table. We joined these networks into one dataset and 100 samples for each binned $\mu$ were randomly selected to plot. The $\Delta_{\mu}$ used for the GLFR generator was set to .3.

    \begin{table}[H]
        \small
        \centering
        \begin{tabular}{|c c c c c c|}
        \hline
            nodes & $\tau_1$ & $\tau_2$ & $\mu$ & <$k$> & $c_{min}$  \\ \hline \hline
            1000 & 2.5 & 1.5 & 0.05 & 14 & 50 \\
            1000 & 2.5 & 1.5 & 0.10 & 14 & 50 \\
            \multicolumn{6}{c}{\smash{\vdots}}\\
            1000 & 2.5 & 1.5 & 0.90 & 14 & 50 \\
            1000 & 2.5 & 1.5 & 0.95 & 14 & 50 \\
            \hline
        \end{tabular}%
        \begin{tabular}{|c c c c c|}
        \hline
            mixing level & parameter & $L_0$ & $L_1$ & $L_2$  \\ \hline \hline
            \multirow{2}{*}{Low} & $\mu^L$ & .33 & .03 & .027  \\
            & $\Delta_\mu^L$ & .03 & .01 & .002  \\
            \multirow{2}{*}{Medium} & $\mu^L$ & .4 & .2 & .1  \\
            & $\Delta_\mu^L$ & .1 & .1 & .1  \\
            \multirow{2}{*}{High} & $\mu^L$ & .8 & .6 & .3  \\
            & $\Delta_\mu^L$ & .2 & .2 & .2  \\
            \hline
        \end{tabular}
        
        \caption{Generator parameters}
        \label{tab:BaseLFRs}
        
    \end{table}

    For this test we compare the result of four algorithms on the HGLFR test set. The first method we use is the Leiden algorithm~\cite{traag2019louvain}. This algorithm is a heuristic method for maximizing generalized modularity, and an improvement on the well known Louvain algorithm~\cite{blondel2008fast}, which was shown in Lu et al.~\cite{lu2020asymptotic} to be unable to handle the resolution limit we inject here. The next method we use is the Multiscale method from Lu et al.~\cite{lu2020asymptotic}, which was designed to handle the resolution limit our benchmark naturally generates. To these two, we add two more algorithms used in the evaluation of GLFR~\cite{le2017glfr} networks. These were the Infomap algorithm~\cite{rosvall2008maps}, an algorithm for finding community structure through random walks in the network and the Label Propagation algorithm~\cite{raghavan2007near}, which propagates labels to nodes based on how frequent they are among neighbors. 

\begin{figure}[H]
        \centering
        \begin{subfigure}[t]{0.5\textwidth}
            \centering
            \includegraphics[width=1\linewidth]{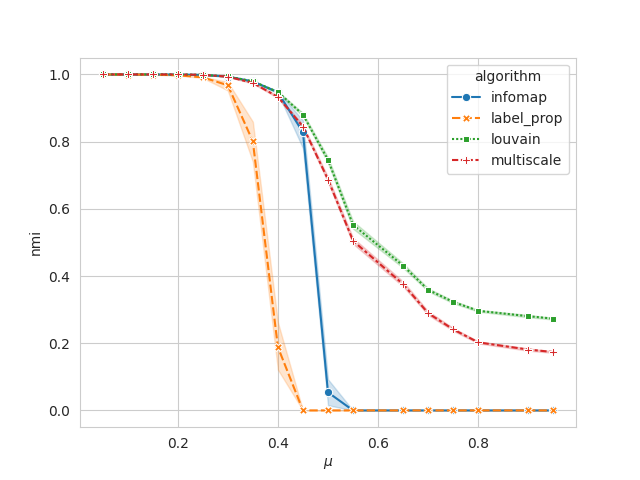}
            \caption{LFR}
            \label{fig:alg_performance_LFR}
        \end{subfigure}%
        \begin{subfigure}[t]{0.5\textwidth}
            \centering
            \includegraphics[width=1\linewidth]{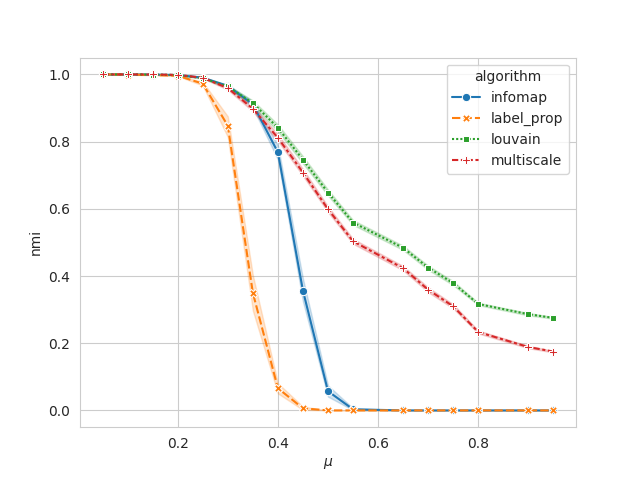}
            \caption{GLFR}
            \label{fig:alg_performance_GLFR}
        \end{subfigure}
        \\
        \begin{subfigure}[t]{0.5\textwidth}
            \centering
            \includegraphics[width=1\linewidth]{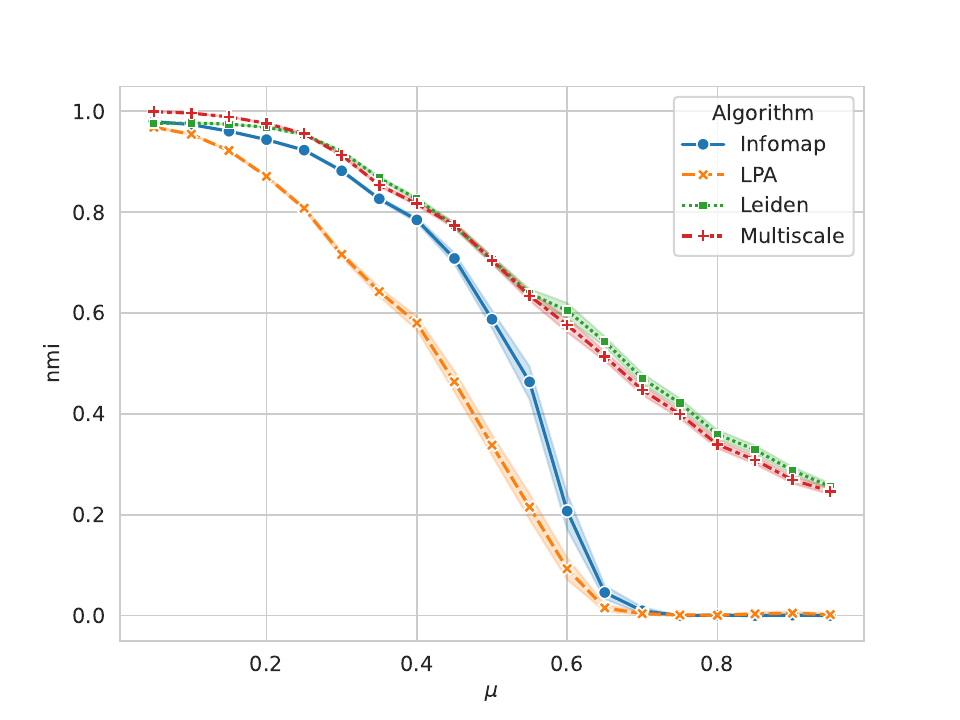}
            \caption{HGLFR}
            \label{fig:alg_performance_HGLFR}
        \end{subfigure}

        \caption{Community detection algorithm performance by $\mu$}
        \label{fig:alg_performance}
    \end{figure}

    Figure \ref{fig:alg_performance} highlights the performance of the tested algorithms against each type of generator. We average each data point over 100 network realizations for the given $\mu$. The performances for all methods are similar between the LFR and GLFR networks. We see that the GLFR generator begins decreasing in performance sooner than the LFR, which is a function of the communities in the network being allowed to vary; thus, networks with a $\mu < .5$ can still contain several communities with $\mu > .5$, thus being more prone to noise and disassortivity.

    The results for the HGLFR networks significantly differ from those of the LFR and GLFR. The low $\mu$ performance is similar to that of the previous two generators but degrades at much lower $\mu$, at around .2. In addition to the quicker drop in performance for all community detection methods; we paradoxically see that the label propagation and Infomap algorithms see less of an overall performance hit for high $\mu (>.5)$.

\begin{figure}[H]
        \centering
            \includegraphics[width=.7\textwidth]{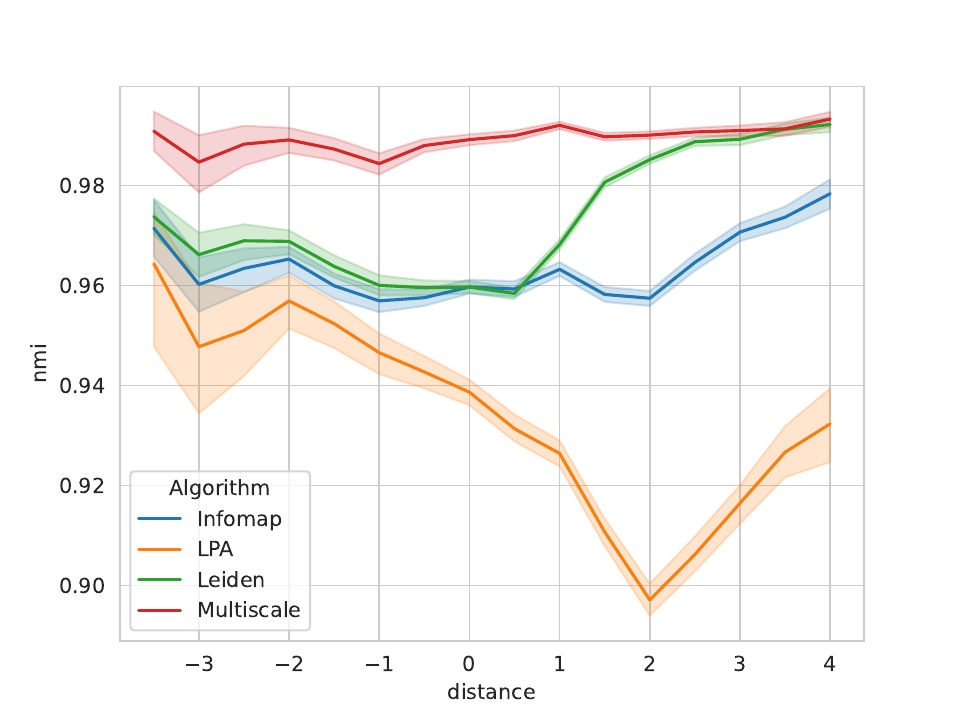}
        \caption{\textbf{Algorithm performance by resolution window distance}. Here we measure the performance of the community detection algorithms against HGLFR networks, varied by $D$. The networks shown all have a $\mu < .2$, so that the negative distances presented are not the result of disassortivity in the ground truth partition. Each point is averaged over 20 network realizations, with the confidence intervals shown around the line.}
        \label{fig:alg_performance_by_distance}
    \end{figure}

    Figure \ref{fig:alg_performance_by_distance} highlights the the benefits of the HGLFR generator when compared to LFR and GLFR. Here we compare the two modularity based community detection methods used previously, this time on networks varied by the resolution window distance $D$. The networks here are all assortative ($\mu < .5$) and the results are averaged over 20 network realizations. For networks with $D < 0$, we see a dip in the performance of the Leiden algorithm, which is improved as the distance increases. The Multiscale algorithm is unaffected by the presence of this resolution limit issue, as it was designed to handle it specifically. These results highlight the strength of this benchmark, which is able to naturally generate networks that are not only more closely aligned with the heterogeneous behaviors seen in real world networks, but is able to model resolution limit issues while maintaining detectable assortative community structures.

    \subsection{Hierarchy detectability}
        We next test the networks generated by the HGLFR generator to validate that the generated hierarchical community structures. To ensure that the generator is creating hierarchical clusters that are high quality, we want to verify that for each level of the hierarchy, that the hierarchical groupings are highly modular for a given range of resolutions.
    
\begin{figure}[H]
            \centering
            \begin{subfigure}[t]{0.5\textwidth}
                \centering
                \includegraphics[width=1\linewidth]{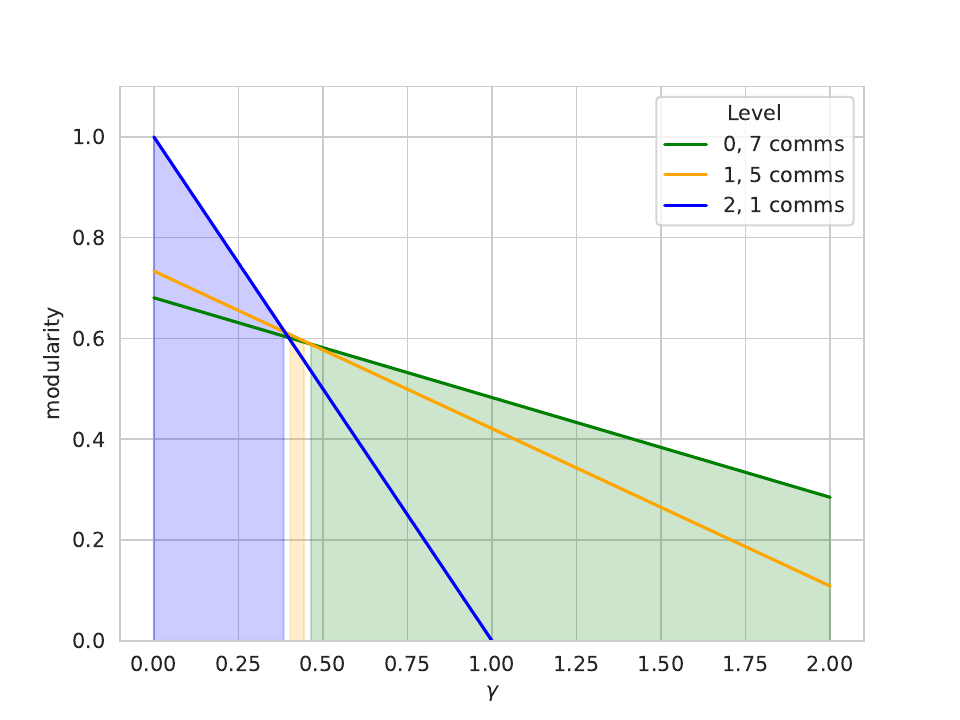}
                \caption{Two Level HGLFR}
                \label{fig:detectability_2_levels}
            \end{subfigure}%
            \begin{subfigure}[t]{0.5\textwidth}
                \centering
                \includegraphics[width=1\linewidth]{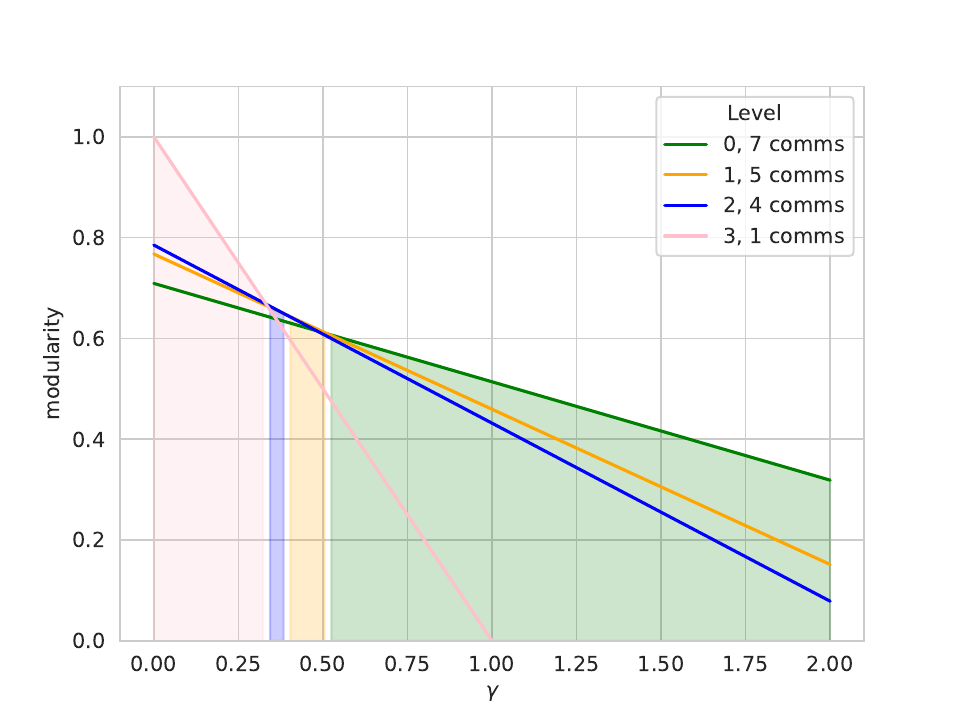}
                \caption{Three Level HGLFR}
                \label{fig:detectability_3_levels}
            \end{subfigure}
            \\
            \begin{subfigure}[t]{0.5\textwidth}
                \centering
                \includegraphics[width=1\linewidth]{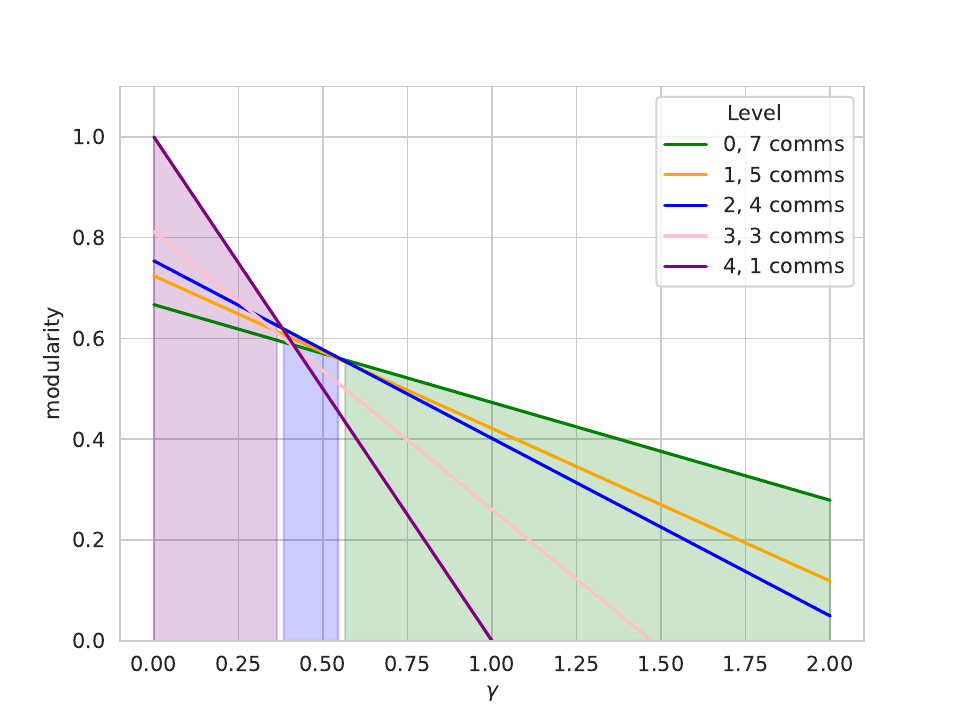}
                \caption{Four Level HGLFR}
                \label{fig:detectability_4_levels}
            \end{subfigure}
    
            \caption{Hierarchical community detectability by resolution ($\gamma$)}
            \label{fig:detectability}
        \end{figure}

        In Figure \ref{fig:detectability} we show, for an example HGLFR network, the convex hull of most modular partitions for a range of given resolution values~\cite{gibson2022finite}. To create this figure, we take each hierarchical level's network partition and compute the generalized modularity of the partition over a large range of resolution values. We would expect that for each hierarchical level, there exists a range of resolutions where the partition is maximally modular, since generalized modularity is designed to find structures of varying coarseness. For this test, we use networks with $D > 0$, since the resolution limit makes detecting the lowest level partition difficult or impossible for generalized modularity methods, per~\cite{lu2020asymptotic}.
        
        Each line represents a partition of the network into one of the hierarchical levels. For lower numbers of levels, like 2 and 3, we can see each level of the hierarchy clearly for a range of resolution values. When we get to larger numbers of hierarchical groupings, we see less levels as having resolutions where they are most modular. In Figure \ref{fig:detectability_4_levels} we see that levels 1 and 3 have no resolutions where they were the highest modularity partition. Despite this, the partitions are still comparably modular, and indicates that the more levels of a hierarchy the harder it is to keep each level strongly distinct.

\section{Conclusion and Future Work}
    
    Community detection benchmarks are an important tool in the validation and testing of community detection methods. Detecting communities in heterogeneous networks is still a challenging problem requiring new approaches. One popular and often efficient approach based on generalized modularity maximization gives rise to anomalies that make the detection of all communities in the network difficult or even impossible for many real networks. To aid in the development of robust improvements and methods to avoid such issues, we introduce the HGLFR network generator, which introduces hierarchical community structure to the widely used LFR benchmark. This new benchmark incorporates further improvements like varied community mixing fractions from other extensions like GLFR and utilizing these changes models real world networks more closely.
    
    More work is needed in the development of robust community detection benchmarks, specifically by carefully evaluating properties of real world networks to model in our benchmark generators. Future avenues of improvement likely lie in the dimension of controlling the degree distributions of the community structure more carefully. Current benchmarks assign nodes, and therefore final degree distributions, to each community randomly. Real world networks may contain targeted variation in how dense different sub-graphs of the network are, and adding heterogeneity to how the community degrees are assigned could solve this. Additionally, further efforts in robust configuration of edges between hierarchical levels is necessary to ensure these benchmark generators sample realizable network space reliably.

\section*{Acknowledgments}
    This work was supported in part by the Army Research Office (ARO) under Grant W911NF-16-1-0524, by the U.S. Department of Homeland Security under Grant Award Number 2017-ST061-CINA01, by the DARPA INCAS Program under Agreement No. HR001121C0165, and the U.S. National Science Foundation under Grant SBE-2214216. The views and conclusions contained in this document are those of the authors and should not be interpreted as representing the official policies either expressed or implied of the Army Research Office, the U.S. Department of Homeland Security, or the U.S. Government.

\bibliography{main}

\end{document}